\documentclass[conference, a4paper]{IEEEtran}
\IEEEoverridecommandlockouts
\usepackage{cite}
\usepackage{amsmath,amssymb,amsfonts}
\usepackage{algorithmic}
\usepackage{graphicx}
\usepackage{textcomp}
\usepackage{xcolor}
\usepackage{csquotes}
\usepackage{subcaption}
\usepackage{booktabs}
\usepackage{multirow}
\usepackage[hidelinks]{hyperref}

\usepackage{siunitx}
\DeclareSIUnit{\belmilliwatt}{Bm}
\DeclareSIUnit{\dBm}{\deci\belmilliwatt}
\DeclareSIUnit{\dBi}{\deci\bel i}

\usepackage[nolist]{acronym} 

\usepackage{xcolor,colortbl}

\newcommand{\highlight}[1]{\noindent\textbf{#1}}

\definecolor{DDGray}{gray}{0.8}
\definecolor{DGray}{gray}{0.82}
\definecolor{Gray}{gray}{0.85}
\definecolor{LGray}{gray}{0.9}
\definecolor{LLGray}{gray}{0.95}

\newcolumntype{e}{>{\columncolor{DDGray}}c}
\newcolumntype{f}{>{\columncolor{DGray}}c}
\newcolumntype{g}{>{\columncolor{Gray}}c}
\newcolumntype{h}{>{\columncolor{LGray}}c}
\newcolumntype{i}{>{\columncolor{LLGray}}c}

\usepackage[inline,shortlabels]{enumitem}

\begin{document}
\bstctlcite{IEEEexample:BSTcontrol}
\title{Measurements of Building Attenuation in 450~MHz LTE Networks\\
}

\author{
	\IEEEauthorblockN{
		Christian Sorgatz\IEEEauthorrefmark{1}\IEEEauthorrefmark{4}, 
		Christian Lüders\IEEEauthorrefmark{3},
		Michael Rademacher\IEEEauthorrefmark{1}\IEEEauthorrefmark{4}  
	}
	\IEEEauthorblockA{\IEEEauthorrefmark{1}Fraunhofer FKIE, Bonn, Germany, {firstname.lastname}@fkie.fraunhofer.de}
	\IEEEauthorblockA{\IEEEauthorrefmark{3}Fachhochschule Südwestfalen, Meschede, Germany, {lueders.christian}@fh-swf.de}
        \IEEEauthorblockA{\IEEEauthorrefmark{4}Hochschule Bonn Rhein-Sieg, Sankt Augustin, Germany, {firstname.lastname}@h-brs.de}
}

\maketitle

\begin{abstract}
This work reports on a measurement study to estimate the attenuation of 450~MHz LTE networks. The \acs{LTE} band 72 is currently deployed in Germany, in particular for smart grid applications. Due to this use-case, we assume that a significant amount of future devices will be deployed stationary and indoor which motivated our campaign. We designed a custom measurement device which uses commercial off-the-shelf hardware to assess the downlink \ac{RSRP} of a public mobile network. In addition, a software has been developed to provide non-experts the possibility to conduct these measurements in the future. This software provides the possibility to determine the indoor position based on ground plans. We conducted measurements at three different buildings. Our results reveal, that the building attenuation of 450~MHz LTE networks is highly heterogeneous and mainly depends on the type of the building, the indoor position and in particular the height of the floor where the device is located.
\end{abstract}


\section{Introduction}\label{sec:introduction}

Mobile networks are a key element for the digitalization of our society. Besides the dominant usage for voice and data on end-user smartphones, current mobile networks play a crucial role in fields like industry 4.0, smart cities and critical infrastructures~\cite{bodenhausen2023challenges}. Well-known examples for the usage in critical infrastructures are intelligent gas, water and electricity meters (so-called smart meters) which are installed in public and private buildings~\cite{avancini2019energy}. As soon as such energy supply networks and in particular their distribution are affected, the requirements for availability and security are particularly high.
A mandatory requirement for coverage within a mobile network at a given quality of service is a sufficient signal level at the \ac{UE}. This is a particular challenge for smart meters which are frequently installed inside buildings and in many cases in the basement.

In 2021, the Federal Network Agency in Germany (\textit{Bundesnetzagentur}) has allocated the \ac{LTE} band 72 (450~MHz) to the \textit{450connect~GmbH} with the mandate of providing a nationwide, highly available \ac{LTE} network for the smart grid and critical infrastructure applications~\cite{bundesnetz}. The network is not accessible by private persons, but is reserved for public service companies and other authorities. It is compliant with the standards of the \ac{3GPP} and supports devices based on \ac{LTE} and \ac{LTE-M}. In the vast majority, the user equipment (i.e.,\ smart meters) will be installed by a (municipal) utility provider.
Compared to the widely used wired network infrastructure~\cite{avancini2019energy}, such a mobile network leads to new practical challenges for utility providers. In particular, it is important for them to understand the challenges for signal attenuation inside modern buildings~\cite{Lora-Insight-Findings}. These challenges need to be incorporated in the planning and roll-out phase for new classes of devices.

In order to increase the planning capabilities for utility providers, we have carried out measurements of the signal strength of a 450~MHz LTE network inside different modern buildings. The contributions of this work are the following:
\begin{itemize}[leftmargin=*]
\item We provide the concept, design and implementation details of a measurement framework to easily assess the downlink \ac{RSRP} of mobile LTE networks. This measurement framework is designed to be used by non-experts, such as employees of a utility provider.
\item We tested the usage of our framework by conducting measurements in three different building types. 
\item We evaluated the results of our measurements in terms of building attenuation, floor height gain and general variation of the \ac{RSRP}.
\item We have uploaded additional material to GitHub, which includes the 3D model of the housing of our sensor, to facilitate similar measurements by the research community\footnote{\url{https://github.com/mclab-hbrs/LTE-Strength-Mapper}}.
\end{itemize}

The rest of this work is structured as follows. In Section \ref{sec:related_work}, we provide background information and refer to related work for building attenuation measurements in mobile networks. Section~\ref{sec:methodology} discusses in detail our methodology providing information about our measurement concept for this work. In particular, we describe the environment~(\ref{subsec:enviroment}), our approach for indoor and outdoor positioning~(\ref{subsec:positioning}), the custom developed hardware~(\ref{subsec:hardware}) and software~(\ref{subsec:software}) and the measurement workflow~(\ref{subsec:procedure}) for the user. Our results are presented in Section~\ref{sec:evaluation} which we summarize in Section~\ref{sec:discussion}.

\section{Background and Related Work}\label{sec:related_work}
Buildings lead to additional losses for radio wave propagation compared to free space. 
Due to the complex structure of buildings and the usage of a variety of materials, the mathematical representation of such losses is a challenging and error-prone task. The \ac{ITU-R} recommendation \mbox{P.2040-1}~\cite{ITU-R-2040-3} provides background information on such a process. In particular, there is a significant dependency on the material of the building, the direction and angle of incidence and the frequency of the radio wave. Additionally, representative parameters for materials are hard to generalize~\cite{ITU-R-2040-3}. Therefore, measurements are used to quantify typical values for different type of building structures. For decades, such measurements have been conducted~\cite{meas-1, meas-3, meas-4, meas-5, meas-7, meas-8}. 

Researchers have adopted a variety of different methodologies to conduct penetration loss measurements. Two exemplary approaches are described in the following. A mixture of both approaches is possible.
\begin{enumerate}[wide]
    \item The first approach \cite{meas-1,meas-3} involves a wideband signal generator as a sender and a spectrum analyzer as a receiver in combination with directional (i.e.\ horn) antennas. The signal generator is placed in close proximity to the building. The directionality of the antennas provides the possibility to individually assess the attenuation of different materials (wood, concrete, windows,~etc.). Additionally, a wide-range of frequencies can be tested in parallel.
    \item The second approach involves the usage of an operational and public \ac{BS} with a fixed position and antenna pattern in combination with standard hardware (i.e.,\ a smartphone with specific software) as receiver. This approach is used in~\cite{meas-4} and provides the advantage that measurements are more efficient and therefore a variety of different buildings can be tested. Disadvantageous is the inflexibility due to the fixed position and frequency range of the \ac{BS}.
\end{enumerate} 

Based on these considerations, we choose the second approach in this contribution.

Emerging from the usage of \textit{mmWave} bands for 5G, propagation modelling for mobile networks (for indoor and outdoor scenarios) has been revisited by various researchers and projects. The work in~\cite{rappaport2017overview} provides a concise overview about these efforts. For the \ac{O2I} penetration loss, two different models are discussed in detail. These models are based on specifications in~\cite{3gpp.36.873,3gpp.38.901} and are continuously updated as part of the \ac{3GPP} releases.
Both models distinguish high loss and low loss buildings.
The decision which model is more applicable for a certain building is not evident. 
As stated in~\cite{3gpp.38.901}, the usage of metal-coated glass in modern buildings and upcoming energy conservation initiatives significantly affect the indoor loss.

Floor height gain is a commonly used term to describe the correlation between the height and the reception power inside the building. As expected, the reception power is greater on a higher floor since the chance for a line-of-sight path between \ac{BS} and \ac{UE} increases. In addition, the amount of blocking buildings decreases. The authors in~\cite{meas-7} measures a floor height gain of  0.6~dB/m in the frequency range below 8~GHz. The authors in~\cite{meas-8} measures a floor height gain of 2~dB/floor for upper floors while the difference between the first and the second floor is significantly greater (6~dB). 

To the best of our knowledge, there is no specialized prior work focusing on 450~MHz LTE networks for indoor signal attenuation. In addition, our proposed framework based on \ac{COTS} hardware has the potential to collect more data compared to previous efforts in this field.

\section{Methodology}\label{sec:methodology}
To quantify building attenuation values, we follow the general idea of recording LTE signal levels inside and outside particular buildings~(\ref{subsec:enviroment}). For this purpose, we have developed software~(\ref{subsec:software}) that reads the necessary values from a \ac{COTS} LTE modem~(\ref{subsec:hardware}), determines a position~(\ref{subsec:positioning}) and stores the combination of both values. The value pairs are displayed on a map and can be exported for further analysis.

In our work, we consider the total attenuation L, from \ac{BS} to the LTE modem, to be of the following simple form: 
\begin{equation}\label{eq:simple-form}
   L = L_{PL} + \chi_{PL} + L_B + \chi_B
\end{equation}
where $L_{PL}$ is the outdoor path loss in combination with a long-term fading random variable ($\chi_{PL}$). $L_B$ is the building attenuation along with a random variable $\chi_B$. This simple form holds, since the distance from the \ac{BS} is significantly larger than the indoor distance.

We can use this formula in combination with our measurement values from outside  ($RXLEV_{o,LM}$) and inside ($RXLEV_{in}$) of the building.

\subsection{Environment}\label{subsec:enviroment}
\begin{figure}
    \centering
    \includegraphics[width=\columnwidth]{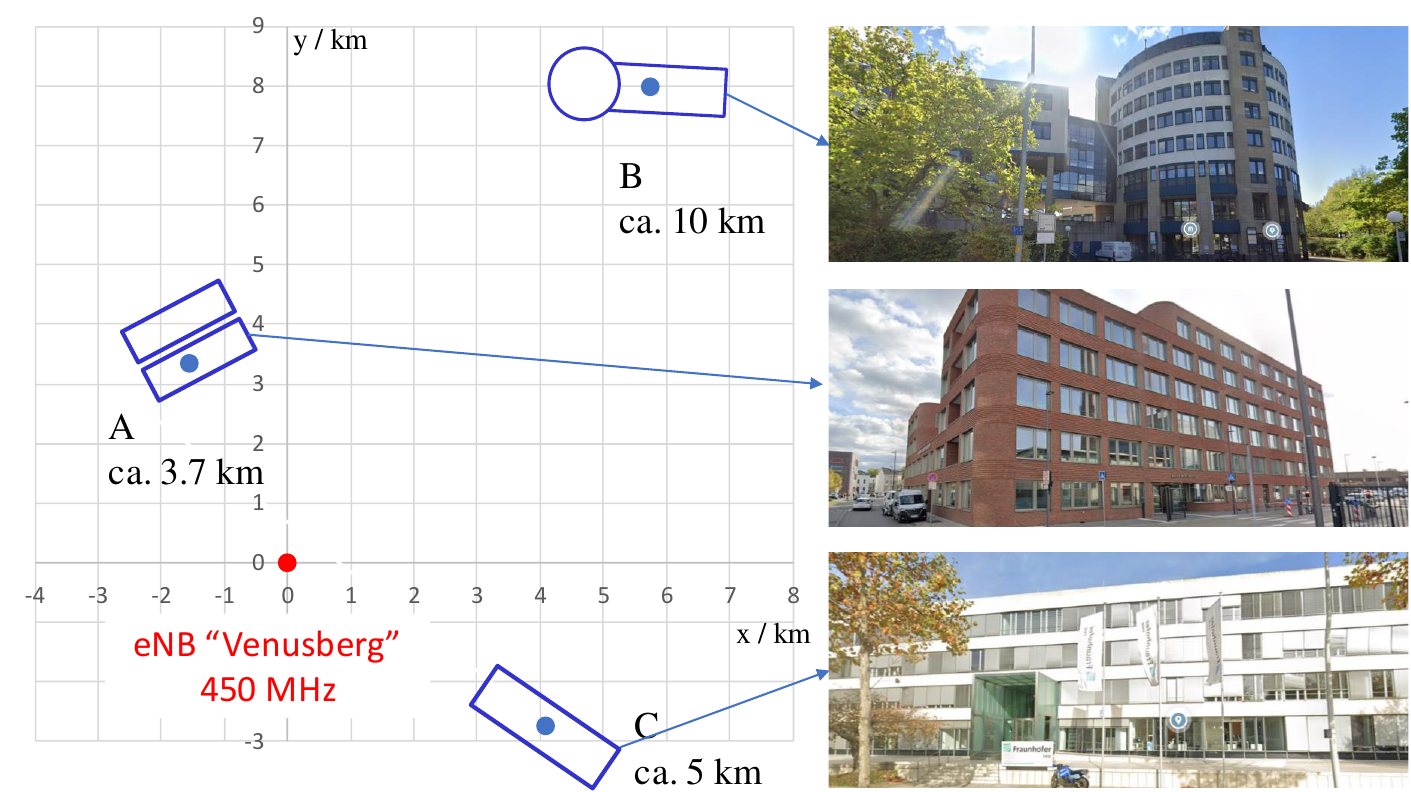}
    \caption{The locations and shapes of the buildings in relation to the \acs{eNB} (left, size of buildings not to scale). A visual impression of the buildings (right).}
    \label{fig:standort}
\end{figure}

The buildings and the radio tower are located in Bonn, Germany.
The city has a population of around 336,000 and is a typical urban environment with medium and tall buildings.
The radio tower is positioned on Bonn's \textit{Venusberg}.
The Venusberg is a high plateau in Bonn, which at 171 m above sea level is around 111 m higher than the center of Bonn~\cite{venusberg}.
However, due to its plateau character, there is no line-of-sight path for different parts of the city.
During the measurement campaign, the 450~MHz network in Bonn is limited to this one \ac{eNB}, which means that all measured signals were transmitted from this tower.

Figure \ref{fig:standort} schematically shows the position and orientation of the buildings in relation to the transmitter in addition to some pictures.

\subsubsection{Building A} %
Building A is a modern office building with two parallel, elongated building sections.
One part of the building has 4 floors above ground and the other has 6 floors.
The lower part of the building is located in front of the higher one as seen from the transmitter.
The building shell is typically thick, and the windows are coated.
The building is about 3.7 km away from the transmitter, but lies in the area of theoretical shadowing by the Venusberg.
\subsubsection{Building B} %
Building B is the furthest building that we have measured.
At a distance of 10 km, the building is no longer located in the city of Bonn.
Due to the elevation on the Venusberg, \ac{LOS} characteristics prevail at the location, which compensates for the path losses due to the greater distance.
The building consists of two connected parts.
One part is an elongated cuboid with one long side facing the transmitter.
The cuboid has 4 floors.
Connected to this cuboid is a cylindrical tower with 7 floors (offices).
Due to hill conditions, the \textit{-1 floor} is only partially underground.

\subsubsection{Building C} %
Building C has 4 levels above the ground.
This building is an office building with coated window panes and solid walls.
It has the shape of an elongated cuboid.
The orientation of the building is unfavorable because the long sides of the cuboid point straight towards the \ac{eNB} on the Venusberg (cf.\ Figure~\ref{fig:standort}).
The building therefore shields the signal itself.
The distance to the \ac{eNB} is about 5 km.

\subsection{Positioning}\label{subsec:positioning}

\acp{GNSS} such as the \ac{GPS} or Galileo are often used to assign the recorded signal levels to a location.
As these systems require the sensor to receive signals from satellites, this type of positioning is suited to outdoor localization.
We therefore only use \ac{GNSS} localization for values recorded outside.
Inside buildings, localization is challenging.
We use the human ability to orientate ourselves in our surroundings.
Equipped with a plan, it is rather simple for a human to determine their current location accurately for our use-case.

There are several options for obtaining site plans of buildings.
The plans can often be provided by the administration.
By providing them in advance, the software can be prepared, which shortens the time required for on-site measurements.
Another option, at least in Germany, is to use the escape and rescue route maps for public buildings, which must be displayed in the corridors.
As these maps are not available in advance, the software must be flexible enough to load the maps on the fly.
For this purpose, it has proven useful to load smartphone photos of the site plans into the software.

\subsection{Hardware}\label{subsec:hardware}
To determine the signal levels of the LTE network at 450~MHz, we use the Quectel BG95-M4 modem, which supports band 72.
The modem is connected to a laptop using a Mini-PCIe to USB adapter and a corresponding extension cable.
The extension cable is used to put distance between the measuring device and the person measuring in order to minimize any interference.
We have developed a housing for the modem that contains an \textit{Arca Swiss} type mount, familiar from photography, in order to attach the modem securely to consumer tripods.
We have made this housing open source, as discussed in Section~\ref{sec:introduction}.
\begin{figure}
    \centering
    \includegraphics[width=\columnwidth]{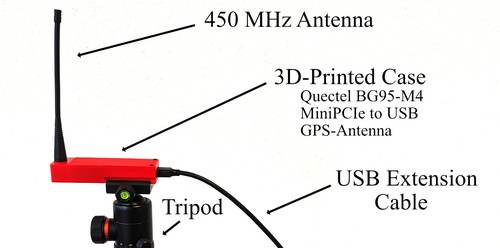}
    \caption{The sensor used for our measurements mounted on consumer tripod at a height of ca. 1.5 m.}
    \label{fig:manual_sensor_v1}
\end{figure}
We were able to obtain the following signal level indicators from the modem: \acf{RSRP}, \ac{RSRQ}, \ac{RSSI}, \ac{SINR}.
For our analysis we use the \ac{RSRP}, since for this parameter, the signal level is averaged over a reference symbol. 
In contrast, the \ac{RSSI} is measured over the entire frequency carrier and therefore also includes possible interferences. This is also evident for the \ac{SINR}. The \ac{RSRQ} is calculated from \ac{RSRP} and \ac{RSSI}~\cite{3gpp.36.214}.

The modem also has a \ac{GNSS} receiver, which can be used for outdoor positioning.
For this purpose, a \ac{GPS}/Galileo antenna is connected to the modem and stowed inside our case.
A $\lambda/4$ antenna is mounted outside the case for the LTE network.
The setup is depicted in Figure~\ref{fig:manual_sensor_v1}.

The modem is connected to a laptop via USB, eliminating the need for an extra power supply, and the software used to record the measured values runs on the same device.

\subsection{Software}\label{subsec:software}

\begin{figure*}[ht]
    \centering
    \begin{subfigure}[b]{0.475\textwidth}
        \includegraphics[width=\columnwidth]{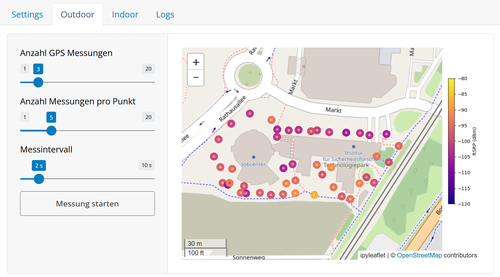}
        \caption{Outdoor Measurement in LTE Strength Mapper}
        \label{fig:mapper_outdoor}
    \end{subfigure}
    \begin{subfigure}[b]{0.475\textwidth}
        \includegraphics[width=\columnwidth]{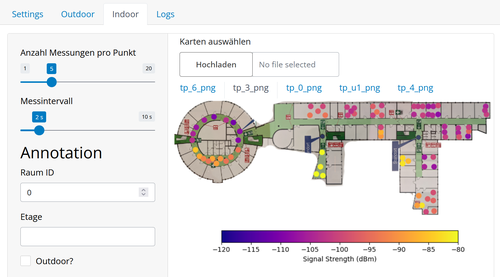}
        \caption{Indoor Measurement in LTE Strength Mapper}
        \label{fig:mapper_indoor}
    \end{subfigure}
    \caption{Developed software framework for our measurement campaign.}
\end{figure*}

We call our software that carries out the measurements and mapping the \textit{LTE Strength Mapper}. 
The software that addresses the modem via a serial interface was implemented in Python.
The \ac{UI} of the software is realized via a web interface.
For this, the package shiny~\cite{shiny} is used, which, among other things, makes interactive Jupyter widgets available in the browser.

For outdoor positioning, we use the \ac{GNSS} receiver provided by the modem.
In the \ac{UI}, the outdoor measurements can be accessed via the \textit{Outdoor} tab.
As shown in Figure~\ref{fig:mapper_outdoor}, this layout consists of a settings column (left) and a visualization column (right).
Three settings can be made here for each measurement.

\begin{itemize}[leftmargin=*]
\item The number of \ac{GNSS} measurements used to determine the position.
In order to reduce the scatter of the positioning, the mean value is determined from all the position determinations.
\item The number of LTE signal level measurements at this position.
The quality of the LTE signal can also fluctuate per position due to propagation effects.
All recorded measured values are saved for further analysis.
\item The measurement interval.
The signal level is measured in idle mode of the modem.
The modem updates the signal level in every \ac{DRX} or \ac{eDRX} cycle.
The value set here should therefore be greater than this cycle to ensure that the modem has already measured a new value.
\end{itemize}

Clicking the button triggers a new measurement.
The following values are saved for each measurement:
\begin{itemize}[leftmargin=*]
\item ID: Each measurement has a specific ID in the form of X.Y where X is the ID of the position and Y is the ID of the measurement on this position.
\item Latitude, longitude and altitude.
The coordinates of the measurements.
\item UTC and date.
The current point in time to match the measurements with a potential protocol.
\item \ac{TAC} and \ac{CID}.
For determining the tower and cell the signal is coming from.
\item The signal level indicators as discussed above.
\end{itemize}

A world map is displayed on the right-hand side, on which the measured point is plotted. The color of the point is an indicator of the quality of the signals received.

The Indoor tab provides the functions for indoor measurement.
As shown in figure \ref{fig:mapper_indoor}, the layout is again divided into two parts.
In the left column, the settings and metadata for the measurement are queried.
In addition to the number of LTE measurements per point and the measurement interval, both of which work in the same way as for outdoor measurements, the following metadata is queried:

\begin{itemize}[leftmargin=*]
\item Room ID.
With the room ID, it is possible to assign several measurements to one room.
For instance, this can be used to determine standard deviations within a room.
\item Floor.
By logging the current floor, analyses can also be carried out per floor and thus the height of the measurement can also be included in the analysis.
\item Outdoor.
This flag is set if the measurement location is outdoors.
For example, if there are plans for an outdoor area, an outdoor measurement can also be carried out.
Balconies, terraces, and roof terraces are typical locations.
\end{itemize}

The following values are saved for each measurement:
\begin{itemize}[leftmargin=*]
\item ID, UTC, date, signal level indicators, \ac{TAC}, \ac{CID}.
As described for the outdoor values.
\item Room, floor and outdoor flag.
This is the metadata as described above.
\item Map. The ID of the floor plan to which the measuring point is assigned.
\item x, y. The coordinates on the image of the floor plan.
This number is dimensionless.
\end{itemize}

The right column of the \ac{UI} offers options for uploading plans, changing the active plan and displaying it.
Plans can be uploaded in pixel image format, making it easy to photograph the escape and rescue plans on premise and upload them to the system.
The plans that are already existing are given their own tab and the ID of the plans is the name of the image file.
The displayed plan is interactive and can be zoomed in and out and panned.
A measurement is started by clicking on the plan.
The position clicked on determines the x, y coordinates of the measurement.
For each measurement carried out, a point appears on the map in a color-coded scale depending on the strength of the LTE network.
If a point is clicked, the values of the measurement are displayed in a pop-up info.

\subsection{Procedure}\label{subsec:procedure}
We will now illustrate a typical measurement sequence as an example.
It is assumed that a suitable building has already been selected and access to the premises has been granted.
In addition, we also assume the site plans are already available.

Before measuring can begin, the software should be prepared with the existing plans.
This involves converting plans in PDF format into an image format (e.g. png) and then renaming the file according to the floor or building section.
Finally, the images are uploaded to the web interface.

For outdoor measurements, it is advisable to measure the signal strength on a pathway completely around the building.
The distance to the building has to be taken into account.
This should not be chosen too low in order to minimize the impact of propagation effects in relation to the building.
Around 2 to 4 meters distance to the building should be sufficient.
The number of measurements required depends on the size of the building and the spread of the values.

For indoor measurements, the ideal scenario would be to take measurements in every room on every floor.
This is rarely practicable in reality, which is why a fair balance should be found.
Once the rooms to be measured have been selected, the tripod with the sensor is moved within the room and the measurement is started at several points.
The number of measurements should be adapted to the size of the room.
For a normal two-person office, five measurement points proved to be enough, as the mean standard deviation within the rooms was rather low with around $3.1$ (Building A), $4.4$ (B) and $2.6$ (C).
Corridors can also be considered as a room, logically separated from other corridor sections.

Once the measurements have been completed, the data saved as a csv-file can be analyzed offline.

\section{Evaluation}\label{sec:evaluation}
In this section, we provide first results based on our measurements in the three different buildings. The results should be regarded as a first indication. The main goal of this work is to introduce our measurement framework as described in the last section. For a detailed assessment of building attenuation in this LTE network more data is needed. In particular, we need more buildings for statistically relevant results. Nevertheless, based on our already conducted measurements, we will provide first indications in the following which shows that our measurement framework is capable of such measurement campaigns.

\begin{figure}[ht!]
    \centering
    \includegraphics[width=\columnwidth]{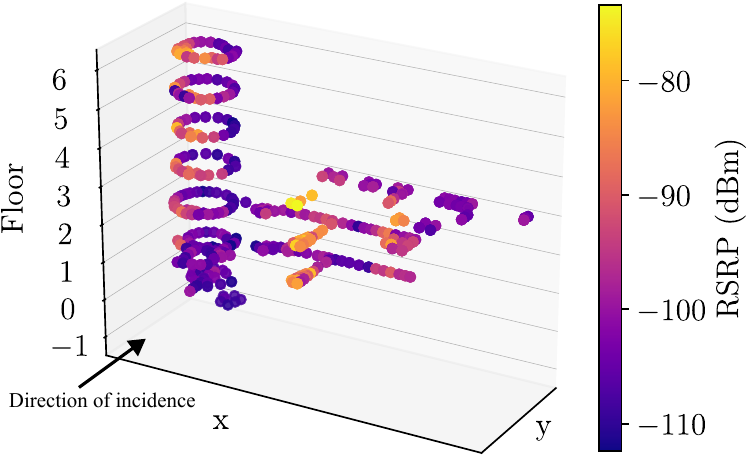}
    \caption{3D scatter plot of the measurements for building B.}
    \label{fig:3dscatter}
\end{figure}

Figure \ref{fig:3dscatter} shows a 3D representation of the data we gathered inside building B. The shape of the building relates to the representation shown in Figure~\ref{fig:standort}. In particular, the cylindrical tower is clearly visible. The signal levels are color-coded and the direction of the \ac{BS} towards the buildings is observable. Overall, a significant difference in the signal level is observable in different dimensions. The difference between the signal level in the direction towards the \ac{BS} and the other side of the building is in particular observable in Figure~\ref{fig:3dscatter}. At the moment, this fact alone leads to a challenge for the utility providers which need an awareness of this direction. This situation may change if the building is served by multiple \ac{BS}. However, since a nation-wide roll-out of the \ac{RAN} infrastructure is challenging for a new network provider, there will likely be time periods where buildings are only served by one cell as shown in our example. 

\begin{figure}[ht!]
    \centering
    \includegraphics[width=1\columnwidth]{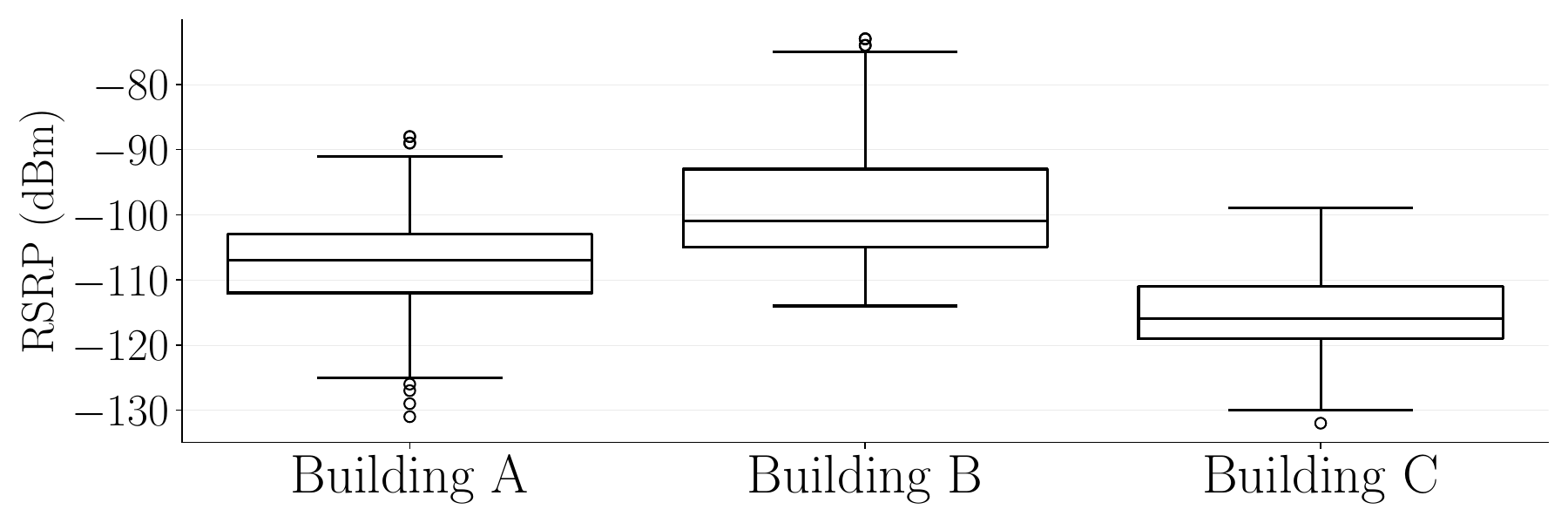}
    \caption{Spread of the \ac{RSRP} values per building. The boxplots are calculated from 255 (A), 1735 (B) and 923 (C) measurements.}
    \label{fig:boxplots}
\end{figure}

The spread of the \ac{RSRP} values is additionally visualized in Figure~\ref{fig:boxplots}, for all measured buildings. For all buildings, the spread is between 40~dB to 50~dB. In addition, in all buildings we measured no reception in the basement at all. 

\begin{figure}[ht!]
    \centering
    \includegraphics[width=1\columnwidth]{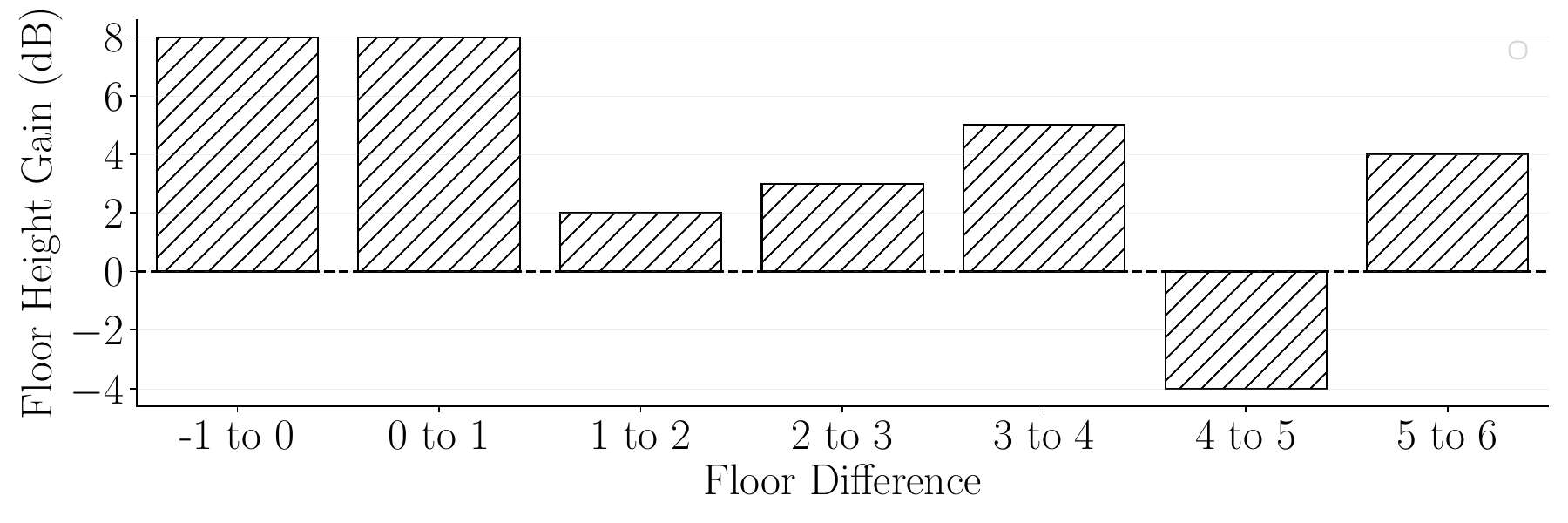}
    \caption{Floor height gain of best \ac{RSRP} value per floor.}
    \label{fig:floor_height_gain}
\end{figure}

As discussed in Section~\ref{sec:related_work}, the floor height gain is an interesting parameter. We visualized this effect for building B in Figure~\ref{fig:floor_height_gain}. The plot shows, that with an increasing floor, the signal level increases as well for 6 out of 7 floors. The floor height gain is in particular evident between the ground floor (-1; as described in \ref{subsec:enviroment}) and the first floor (0) and between the first floor (0) and the second floor (1). Only between the floor 4 and floor 5 there is a decrease in the average signal level. We assume that this effect occurs due to the fact that on floor 5 a different type of office furniture is used. In addition, the office doors are made out of glass compared to wooden doors on the other floors. The overall difference between the basement and upper floor of the cylindrical tower is 26~dB.

Our assumption is that the majority of smart meters will be deployed in the basement. Therefore, in our opinion, a discussion about local network infrastructure is desirable to connect the smart meters in the basement with a possible LTE modem on the upper floors. 

\begin{table}[th!]
\caption{Calculated mean and standard deviations for indoor and outdoor \ac{RSRP}. The indoor \ac{RSRP} values are confined to floors 0 and 1 and the resulting counts are provided.}
\begin{center}
\begin{tabular}{lllll}
\hline
Building                      &       & A      & B      & C      \\ \hline
\multirow{3}{*}{Outdoor RSRP} & Mean [dBm]  & -101   & -98.3  & -101   \\
                              & SD [dB]   & 6.7    & 4.5    & 4.6    \\
                              & Count & 117    & 225    & 306    \\ \hline
\multirow{3}{*}{Indoor RSRP}  & Mean [dBm]  & -122.4 & -101.5 & -116.5 \\
                              & SD [dB]  & 4.2    & 7.2    & 5.6    \\
                              & Count & 22     & 565    & 270    \\ \hline \hline
Building Loss                 & Mean [dBm]  & \textbf{22.4}   & \textbf{3.2}    & \textbf{15.5} \\
(1st/2nd floor)               & SD [dB]   & 7.9    & 8.4    & 7.2    \\

\end{tabular}
\label{tab:building_attenuation_estimation}
\end{center}
\end{table}

Table~\ref{tab:building_attenuation_estimation} is our first approach to assess a typical value for building attenuation in 450~MHz LTE networks. As described in equation~\ref{eq:simple-form}, we simply build the difference between outdoor and indoor measurements. To provide a fair assessment with the floor height gain in mind, we compare the outdoor values on the ground with the indoor values from the 1st and second floor.
Our results show that the building attenuation ($L_B$) for buildings A, B, and C is approximately 22.4~dB, 3.2~dB, and 15.5~dB, respectively, with an uncertainty factor ($\chi_B$) corresponding to the standard deviation.
These results align with the materials of the different buildings. While building B is relatively old, both building A and C are newly built using modern energy saving techniques. This aligns with our building attenuation values.

Again, we want to emphasize that we were unable to obtain any signal level greater than the minimum sensitivity of our modem (-140~dBm) in the basements of the buildings. Therefore, the building attenuation towards the basements is greater than 40~dB for all three different buildings.

\section{Summary and Future Work}\label{sec:discussion}
This work presents a measurement framework for building attenuation in LTE networks. We successfully used this framework to obtain data in a 450~MHz LTE-M network from three different buildings and analyzed the data based on different criteria. The results reveal that the signal level throughout the same building is highly heterogeneous. There is a huge spread between 40~db to 50~dB depending on the type of building. A significant parameter is the height. The reception is significantly better on the upper floors. 

In the current state of the network deployment, in all three buildings, we were unable to measure a signal level in the basement at all. One of the ideas of this new LTE network is to advance the digitalization of energy networks such gas, water or electricity, however, the relevant measurements points for such networks are typically located in a basement in Germany.

Our main focus for future work is to use our framework and obtain more data from different buildings. This is an important step for a statistically sound evaluation of building attenuation. Due to the modular nature of our framework, we are not limited to measure LTE but could easily extend to 5G networks by simply changing the hardware modem. The general concept and the majority of software can be reused. 

\highlight{Acknowledgment}
This work has been funded by the German Federal Office for Information Security (\ac{BSI}) under project funding reference number 01MO23003B (PlusMoSmart).
The responsibility for the content of this publication lies with the authors.

\begin{acronym}[]

\acro{3GPP}{3rd Generation Partnership Project}
\acro{BSI}{Bundesamt für Sicherheit in der Informationstechnik}
\acro{BS}{Base Station}
\acro{COTS}{Commercial Off-the-Shelf}
\acro{eNB}{eNodeB}
\acro{GNSS}{Global Navigation Satellite System}
\acro{UE}{User Equipment}
\acro{LTE}{Long Term Evolution}
\acro{LTE-M}{\ac{LTE} Machine Type Communication}
\acro{RSRP}{Reference Signal Received Power}
\acro{RSRQ}{Reference Signal Received Quality}
\acro{RSSI}{Received Signal Strength Indication}
\acro{O2I}{Outdoor-to-Indoor}
\acro{RSSI}{Received Signal Strength Indication}
\acro{TAC}{Tracking Area Code}
\acro{GPS}{Global Positioning System}
\acro{RAN}{Radio Access Network}
\acro{UI}{User Interface}
\acro{ITU-R}{\ac{ITU} Radiocommunication Sector}
\acro{ITU}{International Telecommunication Union}
\acro{LOS}{Line of Sight}
\acro{SINR}{Signal-to-Noise-plus-Interference Ratio}
\acro{DRX}{Discontinuous Reception}
\acro{eDRX}{Extended Discontinuous Reception}
\acro{CID}{Cell ID}
\end{acronym}

\bibliographystyle{IEEEtran}
\bibliography{./bibabbrv.bib, ./lit.bib}
\end{document}